# Recognition of Fermi-arc states through the magnetoresistance quantum oscillations in Dirac semimetal $Cd_3As_2$ nanoplates

Guolin Zheng[1*], Min Wu[1,2*], Hongwei Zhang[1,2*], Weiwei Chu[1,2], Wenshuai Gao[1,2], Jianwen Lu[1,2], Yuyan Han[1], Jiyong Yang[1], Haifeng Du[1,3], Wei Ning[1 §], Yuheng Zhang[1] and Mingliang Tian[1,3,4 §]

[1]*Anhui Province Key Laboratory of Condensed Matter Physics at Extreme Conditions, High Magnetic Field Laboratory, Chinese Academy of Sciences, Hefei 230031, P. R. China.*

[2]*Department of physics, University of Science and Technology of China, Hefei 230026, P. R. China.*

[3]*Department of Physics, school of Physics and Materials Science, Anhui University, Hefei 230601, Anhui, P. R. China.*

[4]*Collaborative Innovation Center of Advanced Microstructures, Nanjing University, Nanjing 210093, P. R. China.*

[*] Those authors contribute equally to this work.

[§]To whom correspondence should be addressed. E-mail: ningwei@hmfl.ac.cn (W. N.), tianml@hmfl.ac.cn (M. T.).

# Abstract

The disjointed Fermi-arcs in Weyl semimetals can intertwine with chiral bulk modes and participate in unusual closed magnetic orbits in the presence of a vertical magnetic field. Here we carry out the quantum oscillation study of such unusual Weyl magnetic orbits in Dirac semimetal $Cd_3As_2$, a close cousin of Weyl semimetals. We find that extra two-dimensional (2D) quantum oscillations emerge at high field, which superimpose on 3D bulk background and can be attributed to the Weyl magnetic orbits, when the thickness of nanoplates is smaller than the mean free path of the electrons. Further evidence of the 2D quantum oscillations from the Weyl magnetic orbits is provided by the nonlocal detection, which demonstrates an alternative way to study the quantum transport properties of Fermi-arcs under magnetic field.

Weyl semimetals (WSMs) are newly emergent topological states of matter whose low-energy bulk excitations can be described by the Weyl equations [1]. These low-energy quasiparticles behave as “Weyl Fermions”, well known in high-energy physics, and exhibit an apparent violation of charge conservation, leading to many exotic transport phenomena, such as chiral anomaly [2-8] and anomalous Hall effect [9-12]. Another signature of the nontrivial topological nature of WSMs is the Fermi-arc surface state, which connects pairs of Weyl nodes with opposite chiralities [13-20]. The disjointed Fermi-arcs can in principle be detected via the quantum oscillatory experiments, if the electrons keep phase coherence after the periodic motion along the Weyl magnetic orbits traversing the bulk to connect the Fermi-arcs on opposite surfaces [21]. Generally, the quantum oscillations from surface Fermi-arcs are mixed with bulk states, leading to a beating pattern [22]. The surface state contribution can only be identified in frequency domain. The spin polarized Fermi-arcs behave as helical edge states, supporting for a nonlocal detection which may extract the surface quantum oscillations from large bulk background and provide other transport signatures of Weyl magnetic orbits.

In this paper, we carry out the quantum oscillation study of the Weyl magnetic orbits in Dirac semimetal $Cd_3As_2$, a close cousin of WSMs. We find that our $Cd_3As_2$ nanoplates exhibit extra weak 2D oscillations superimposed on 3D bulk oscillations at high field, when the thickness of $Cd_3As_2$ nanoplates is smaller than the quantum mean free path of the electrons. The analysis of the oscillation frequency reveals that the 2D oscillations at high field can be attributed to the Weyl magnetic orbits connecting the Fermi-arcs from opposite surfaces. Further evidence for the 2D oscillations of Weyl magnetic orbits is provided by the nonlocal detection, which demonstrates an alternative way to study the Fermi-arcs in WSMs.

$Cd_3As_2$ nanoplates were grown by chemical vapor deposition (CVD) methods on silicon substrates, as discussed elsewhere [4, 5]. To improve the quality of $Cd_3As_2$ nanoplates, the quartz tube was cleaned by ultrapure Ar gas for five times and a background vacuum as low as 15 $\mathrm{mtorr}$ was obtained prior to the growth. The $Cd_3As_2$ nanoplates of length up to 50 $\mu\mathrm{m}$ can be obtained, as shown in Fig. 1(a). $Cd_3As_2$-based quantum devices were fabricated by standard electron-beam lithography (EBL) followed by Au (100 nm)/Ti (10 nm) evaporation and lift-off process.

In Dirac semimetal $Cd_3As_2$, two sets of Weyl points with opposite chiralities are superimposed along $k_z$ direction, protected by the extra crystallographic symmetry from mixing and annihilation [14]. Consider a single pair of bulk Weyl nodes with chiralities $\pm 1$, the Fermi-arcs from two opposite real-space surfaces can participate in a Weyl magnetic orbit under a vertical magnetic field [21], as illustrated in Fig. 1(b). Typically, the Shubnikov–de Haas (SdH) oscillations in these $Cd_3As_2$ nanoplates exhibit a single frequency in low field region but show an extra oscillatory frequency above 5 T, as we can see in Fig. 1(c). This can also be seen in Fast Fourier Transformation (FFT), two oscillatory frequencies with $F_1 = 17.3\ \mathrm{T}$ and $F_2 = 47.6\ \mathrm{T}$ can be easily identified in sample S1. Bulk crystal $Cd_3As_2$ usually presents single-frequency quantum oscillations due to the near-isotropic Fermi surface [23-24].

The extra oscillation frequency at high field is probably due to the quantum oscillation from Fermi-arcs. As aforementioned, a prerequisite for the observation of the Fermi-arc oscillations is that the electrons should keep phase coherence after the periodic motion along the Weyl magnetic orbits. That is the Fermi-arc oscillations can only be observed in thin $Cd_3As_2$ nanoplates with a magnetic field normal to the surface. Rotating the field B from out-of-plane to in-plane direction, the surface Fermi-arc signals will disappear gradually, due to the losing of phase coherence. The observed Fermi-arc oscillations will only depend on the vertical field component, and thus lead to a 2D oscillation pattern. Such a 2D oscillation pattern in 3D Dirac semimetal $Cd_3As_2$ nanoplates is a transport signature of surface Fermi-arcs, which can be testified by analyzing the angular-dependent quantum oscillations.

To testify the 2D Fermi-arc oscillations, we now analyze the angle-dependent magnetoresistance in sample S2. Fig. 2(a) shows the scanning electron microscopy (SEM) image of sample S2. For magnetic field B oriented perpendicular to the surface ($\theta = 0^{\circ}$), the longitudinal magnetoresistance ($R_{xx} = V_{54}/I_{12}$) of S2 shows apparent multi-periodic oscillations. While in in-plane direction with B parallel to the current $I$ ($\theta = 90^{\circ}$), single-frequency oscillations are revealed, as shown in Fig. 2(b). A plot of oscillatory components $\Delta R_{xx}$ versus $1/B$ under various tilted angle $\theta$ is presented in Fig. 2(c). Similar to sample S1, S2 shows single-frequency oscillations (bulk state) in low field region and the peak positions remain unchanged for various tilted angle $\theta$, consistent with the near-isotropic band structure in $Cd_3As_2$. Above 8 T, extra oscillations appear sequentially and the oscillations from bulk state seem to be overwhelmed near $\theta = 0^{\circ}$, as indicated by the blue arrow in Fig. 2(c). As $\theta$ increases, the newly emergent oscillations at high field are shrunk and vanished eventually near $\theta = 90^{\circ}$. If we extract the perpendicular components of these quantum oscillations, both 3D (grey arrow) and 2D characters (grey lines) are revealed, as we can see in Fig. 2(d).

For further identification of the 2D Fermi-arc oscillations at high field, we carry out extra measurements in sample S3, where the magnetic field is randomly rotated from the out-of-plane direction to the in-plane direction, as shown in Fig. 3. Similar situation has also been observed in S3, where extra oscillations emerge at high field, as indicated by the blue arrow in Fig. 3(a). Those extra oscillations at high field also exhibit a 2D character, as indicated by the grey lines in 3(b). As mentioned above, Fermi-arcs can intertwine with the chiral bulk modes and participate in closed magnetic orbits, forming a 2D oscillation pattern. The observed extra 2D oscillations at high field are most probably due to the Fermi-arc oscillations associated with the unusual Weyl magnetic orbits. Now, we theoretically analyze the possibility of observation of the Fermi-arc states in these $Cd_3As_2$ nanoplates. According to the recent theory, the quantum oscillation frequency from surface Fermi-arcs can be estimated by the formula [21]:

$$F_s = E_F k_0/(e\pi v_F), \quad (1)$$

where $E_F$ is Fermi level, $k_0$ is the arc length which is comparable with the separation of the Dirac nodes ($\sim 0.1\ \text{Å}^{-1}$) [21, 23], and $v_F$ is Fermi velocity. For simplicity and without loss of generality, we now calculate the Fermi energy and Fermi velocity in sample S3 via the low-field quantum oscillations. According to the Lifshitz-Kosevich formula, the oscillation amplitudes can be written as [26]

$$\frac{\Delta\rho(T,B)}{\rho(B=0)} \propto \left(\frac{\hbar\omega_c}{E_F}\right)^{\frac{1}{2}} \frac{\alpha T m^*/B}{\sinh[\alpha T m^*/B]} \exp(-\alpha T_D m^*/B). \quad (2)$$

Where $T_D$ is the Dingle temperature and $\alpha = 2\pi^2 k_B/\hbar e$, $k_B$ is the Boltzmann constant. The cyclotron frequency has $\omega_c = eB/m^*$, with $m^*$ the effective cyclotron mass. The Fermi energy $E_F$ can be estimated by $E_F = m^* v_F^2$. Fig. 3(c) shows the temperature dependence of the oscillation components $\Delta R_{xx}$ versus $1/B$ with magnetic field oriented perpendicular to the surface ($\varphi = 0^\circ$). By fitting the oscillation amplitudes to the Lifshitz-Kosevich formula, we get the effective mass $m^* = 0.035\, m_e$, as shown in Fig. 3(d). According to the Onsager relation, $F = (\hbar/2\pi e)S_F$, where $F$ is the bulk oscillation frequency and $S_F$ is the cross section of Fermi surface, we can further calculate the Fermi wave vector $k_F$ which is $0.031\,\text{Å}^{-1}$. Finally, the estimated Fermi energy is 210 $meV$ and Fermi velocity is $1.02 \times 10^6\, m/s$, the calculated Fermi-arc oscillation frequency is $F_s \approx 66$ T. In our observation, the newly emergent oscillations at high field have a spacing $\Delta(1/B) \approx 0.0132\,\mathrm{T}^{-1}$, generating an oscillation frequency 75 T, which is quantitatively consistent with the theoretical calculation. To keep the phase coherence of the electrons after a periodic motion, the quantum mean free length $l_Q$ should be larger than the thickness of $Cd_3As_2$ nanoplates L. The quantum mean free path is given by $l_Q = v_F \tau_Q$, where scattering time $\tau_Q$ has

$$\tau_Q = \hbar/(2\pi k_B T_D). \quad (3)$$

As shown in the inset of Fig. 3(d), the Dingle temperature $T_D$ derived from the slope of the Dingle plot is 13.8 K. Thus, the calculated $l_Q$ in S3 is about 90 nm, which is larger than the thickness of S3 (~80 nm). Thus the Fermi-arc states, appearing as the extra 2D oscillations in high field region, are confirmed combining our experimental observations and previous discussions.

Fig. 4 shows the both local ($R_l = V_{54}/I_{12}$) and nonlocal ($R_{nl} = V_{56}/I_{43}$) magnetoresistance oscillation components of sample S2. Interestingly, the testified Fermi-arc oscillations in local detection can be significantly manifested in nonlocal configuration, as indicated by the grey lines in Fig. 4(a) at $\theta = 0^{\circ}$. Moreover, the bulk states, behaving as local magnetoresistance oscillations at low field, have been largely suppressed here in nonlocal configuration. Thus the nonlocal detection may provide an effective way to extract the Fermi-arc oscillations from large bulk background. If we track the nonlocal oscillations under different tilted angle $\theta$, a more obvious 2D oscillation pattern can be obtained, as shown in Fig. 4(b). The bulk transport in conventional metal, according to the conventional diffusive electronics, satisfies Ohm's law and behaves as local resistivity or conductive tensor [27]. However, in nonlocal configuration, the bulk signal attenuates exponentially as the increasing of the separation $b$ (inset of Fig. 4(a)), according to the van der Pauw formalism [28]. In WSMs, the surface state as an open and disjointed Fermi-arc provides a unique helical conductive path allowing for a nonlocal detection, similar to the edge state in quantum spin Hall or quantum Hall system [27]. Whereas, different from the quantum spin Hall insulator state or 3D topological insulator with gapped and insulating bulk states, the gapless bulk states in WSMs couple with surface

Fermi-arcs, leading to the quasiparticles scattering of surface Fermi-arcs into the bulk [29]. The observed large nonlocal 2D Fermi-arc oscillations indicate that, in the presence of a vertical magnetic field, the surface quasiparticles driven by Lorentz-like force can consecutively slide along the Fermi-arcs and form conductive helical edge states, supporting for a nonlocal detection. We believe that the quasiparticles scattering of surface Fermi-arcs into the bulk can be effectively suppressed by the magnetic field. The vanishing bulk signals demonstrate that non-locality provides an alternative way to study the transport properties of Fermi-arc states in Dirac/Weyl semimetal systems.

Thus far, we have only considered a single pair of Weyl points with opposite chiralities. While Dirac semimetal $Cd_3As_2$ has two copies of superimposed Weyl points along $k_z$ direction, forming the double Fermi-arc surface states [14]. Recently, Kargarian *et al.* [30] pointed out that in the presence of a surface perturbation, the double Fermi-arcs are unstable and can be continuously deformed into a closed Fermi contour. This surface perturbation can theoretically be introduced by magnetic field, which reduces the crystal symmetry [21, 31]. The observed surface state oscillations in high field may or may not origin from the unusual magnetic orbits, depending on the height of Fermi level. In our $Cd_3As_2$ nanoplates, the Fermi level is about 200 $meV$ above the Dirac point. The surface Fermi pockets might merge into the bulk state, forming a similar magnetic orbit, as discussed in Ref. 30. Whereas, there are no surface state oscillations in low field region in our samples, the initially observed 2D surface oscillations should be attributed to the magnetic orbits discussed in Ref. 21. If Fermi-arcs are deformed continuously by the magnetic field, a corresponding descent of the surface oscillation frequency should be observed. This seems not to be the case of our observations, because the surface oscillation frequency keeps unchanged as field $B$ increases. Extra nonlocal experiments are on-going to investigate the deformation of the Fermi-arc states under more intense magnetic field.

## Acknowledgments

We thank professor E. V. Gorbar and professor Kun Yang for fruitful discussion. This work was supported by the National Key Research and Development Program of China No.2016YFA0401003, 2017YFA0303201; the Natural Science Foundation of China (Grant No.11374302, No.U1432251).

## Figure Captions

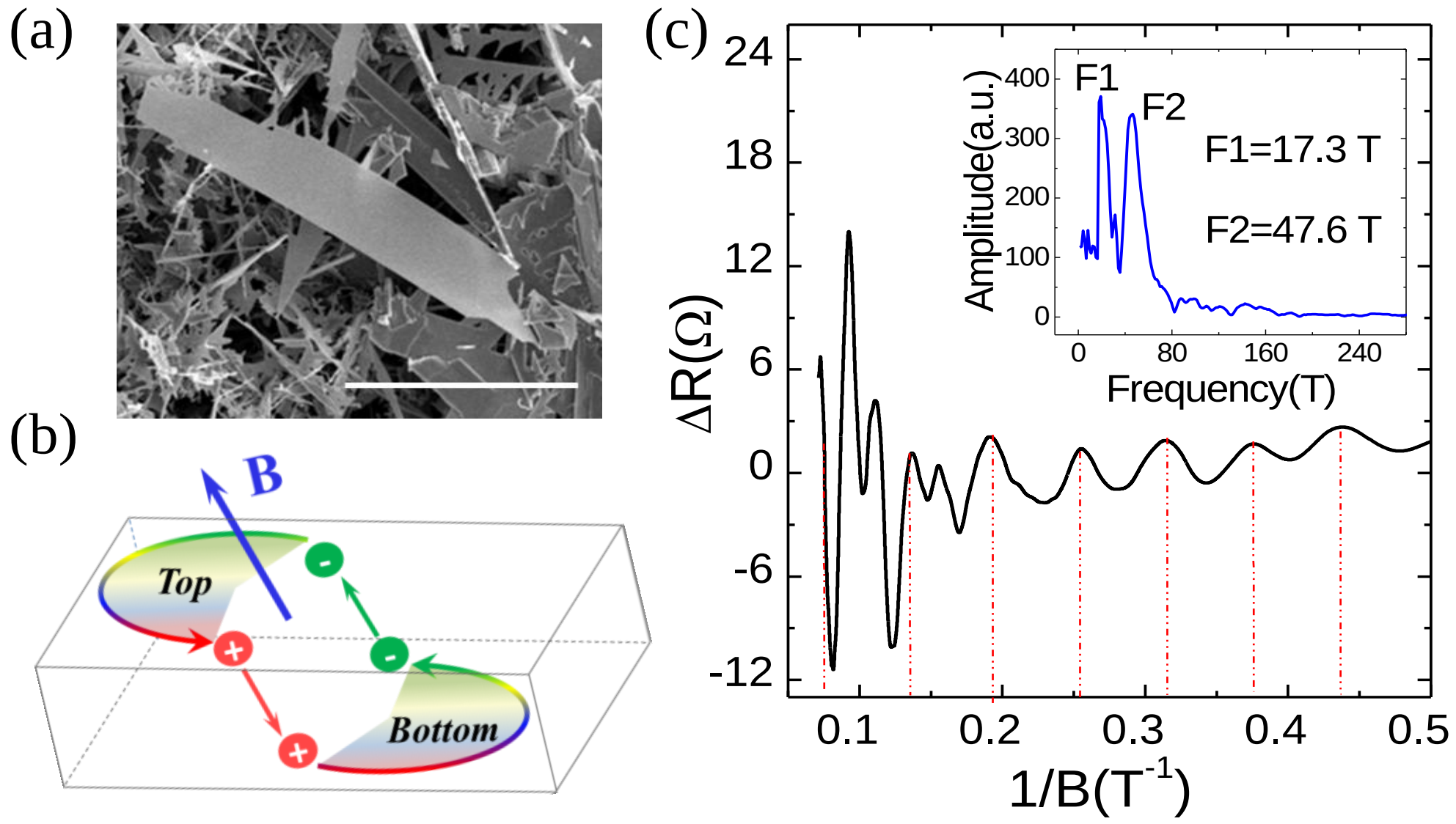

FIG. 1 (Color online). (a) The scanning electron microscopy (SEM) image of the $Cd_3As_2$ nanoplates. Scale bar: 25 μm. (b) An illustration of the Weyl magnetic orbit in WSMs in the presence of a static magnetic field oriented perpendicular to the two opposite surfaces. (c) SdH oscillation components of longitudinal magnetoresistance in sample S1. Extra oscillations emerge at high field. Inset: the corresponding FFT spectra of SdH oscillations.

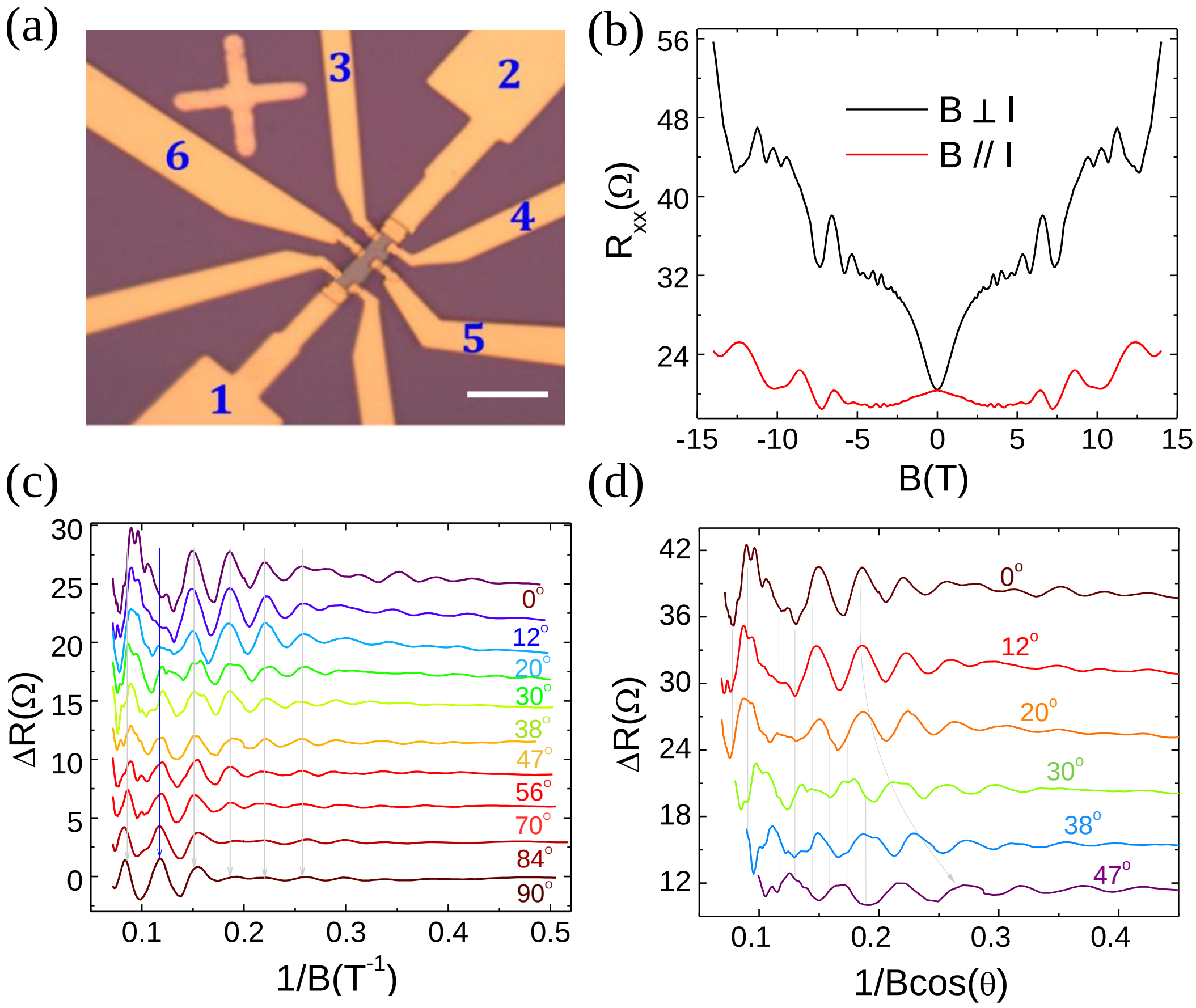

FIG. 2 (Color online). (a) The SEM image of sample S2. Scale bar: 10 μm. (b) The longitudinal Magnetoresistance of sample S2 with $B$ oriented perpendicular (black) and parallel (red) to the surface. (c) Detailed SdH oscillation components under different tilted angle $\theta$, after subtracting the smooth background of MR curves. Extra oscillations emerged at high field have largely overwhelmed the bulk states near $\theta = 0^{\circ}$, as indicated by the blue arrow. (d) Angular dependence of the quantum oscillations. Both 2D character (grey lines) for surface Fermi-arcs and 3D (grey arrow) for bulk states are revealed. Note that, some valleys of 2D surface oscillations are superimposed on the peaks of 3D bulk oscillations.

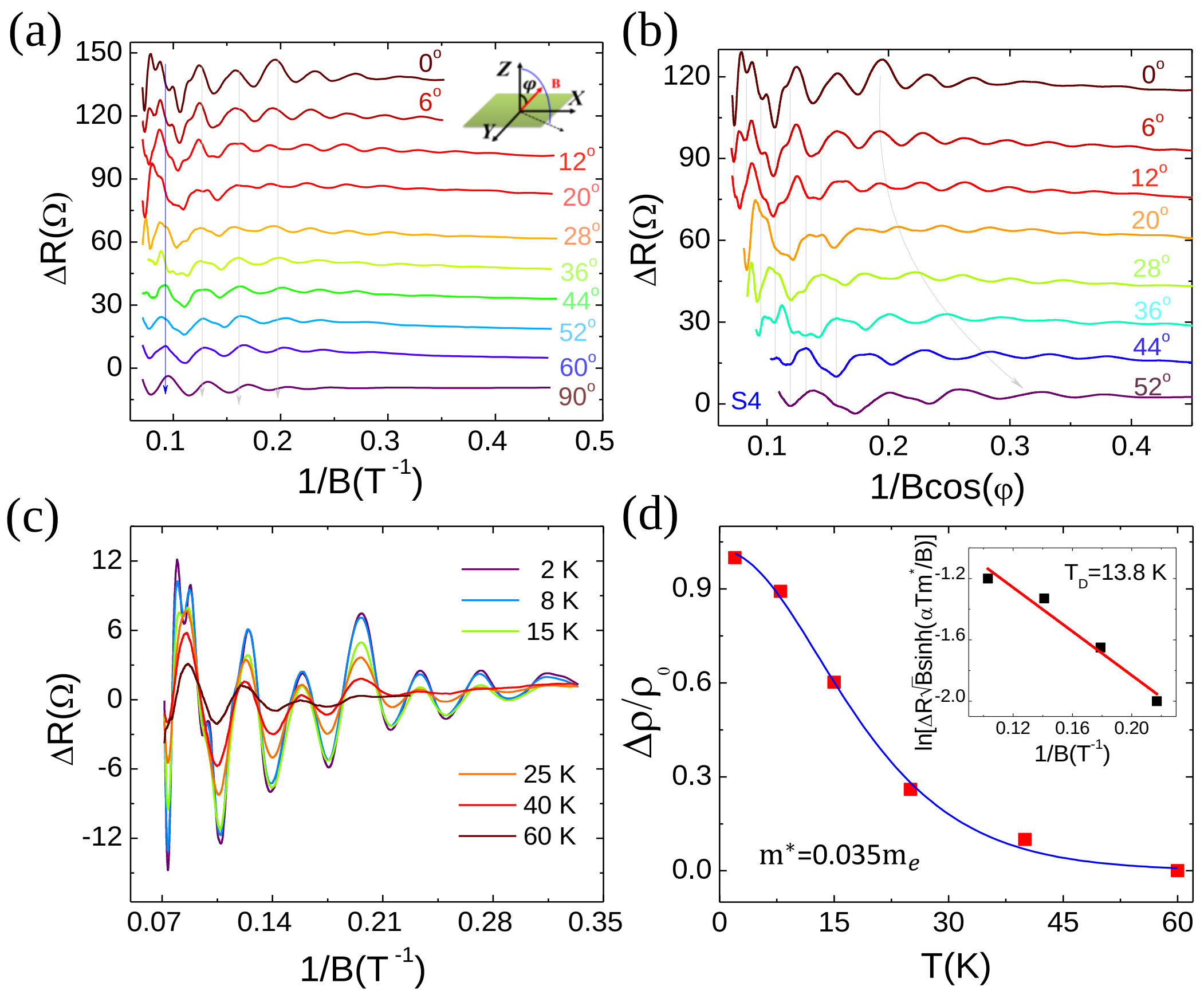

FIG. 3 (Color online). (a) The angular-dependent MR oscillation components in sample S3 (~80 nm) with $B$ tilted randomly from out-of-plane direction ( $\varphi = 0^{\circ}$) to in-plane direction ( $\varphi = 90^{\circ}$). Current is applied along X-direction. (b) The SdH oscillation components versus $1/Bcos(\varphi)$ under different tilted angle $\varphi$. (c) The oscillation components as the function of $1/B$ for $\varphi = 0^{\circ}$ under various temperatures. (d) Fitting the temperature-dependent oscillation amplitudes $\Delta\rho/\rho_0$ $(1/B = 0.242)$ generates an effective mass $m^* = 0.035m_e$. Inset: The fitted Dingle plot at 2 K generates a Dingle temperature $T_D = 13.8$ K.

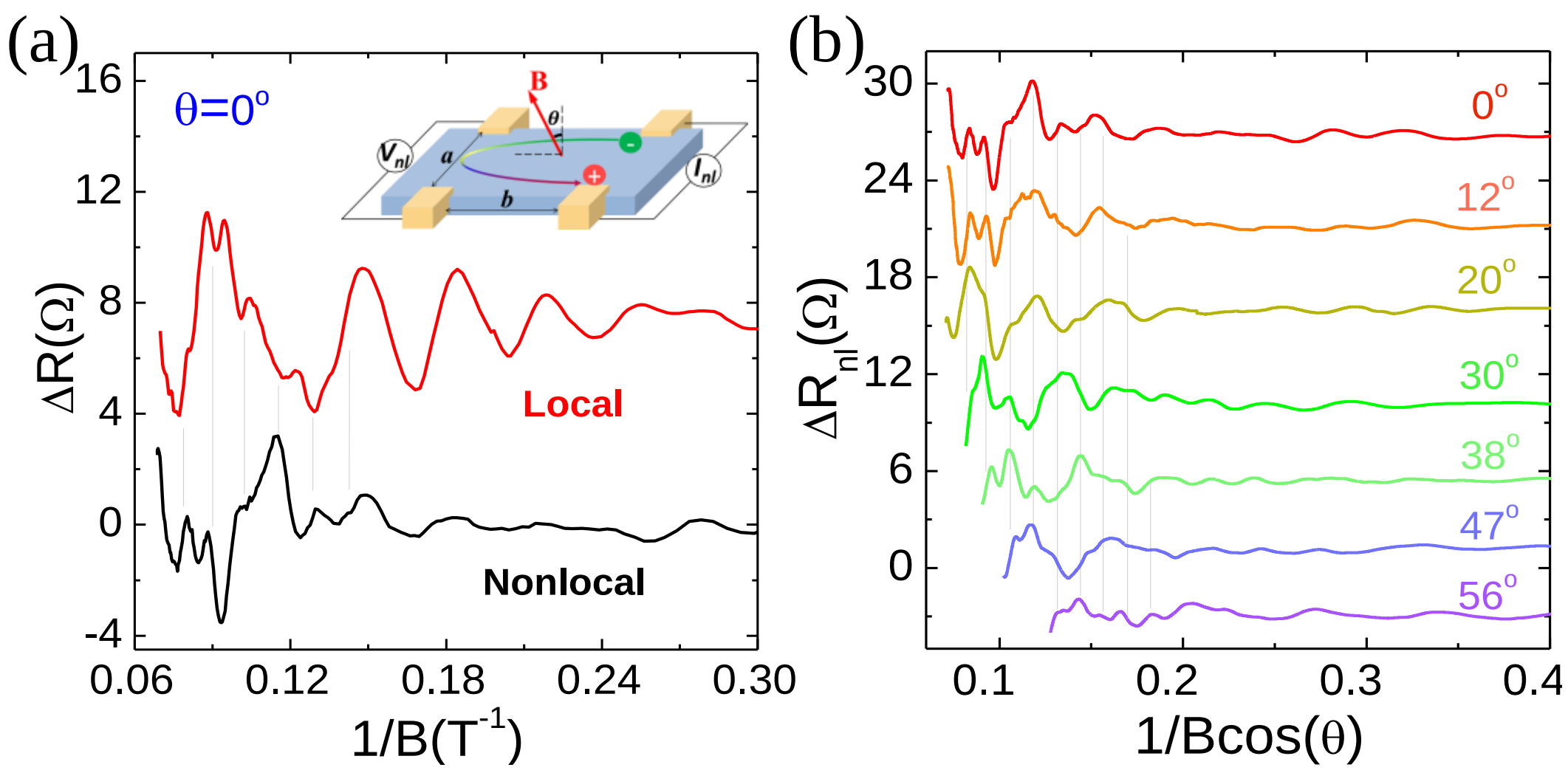

FIG. 4 (Color online). (a) A comparison of the local ( $R_l = V_{54}/I_{12}$ ) and nonlocal ( $R_{nl} = V_{56}/I_{43}$ ) oscillation components in sample S2 for $\theta = 0^\circ$. The 2D Fermi-arc oscillations in local detection can be significantly manifested in nonlocal configuration with vanishing bulk contribution. Inset: a schematic illustration of the nonlocal detection. In sample S2, $b/a \sim 2.7$. (b) A nonlocal manifestation of the 2D Fermi-arc oscillations.